\documentstyle[11pt,twoside,epsfig,A4]{article}

\begin{document}

\newcommand{\beg}{\begin{equation}}
\newcommand{\eeg}{\end{equation}}
\newcommand{\br}{\begin{array}}
\newcommand{\er}{\end{array}}
\newcommand{\bea}{\begin{equation} \begin{array}{c}}
\newcommand{\eea}{\end{array} \end{equation}}
\newcommand{\un}{\underline}
\newcommand{\cb}{\begin{center}}
\newcommand{\ce}{\end{center}}
\newcommand{\llg}{\left\langle}
\newcommand{\rrg}{\right\rangle}
\newcommand{\thh}{$^{\mbox{th}}$}
\newcommand{\al}{\alpha}
\newcommand{\bite}{\begin{itemize}}
\newcommand{\eite}{\end{itemize}}
\newcommand{\m}{\mbox}
\newcommand{\wh}{\hspace{3mm} \mbox{ where }}
\newcommand{\lcm}{\mbox{lcm}}
\newcommand{\di}{\mbox{ div }}
\newcommand{\T}{\mbox{ TRUE }}
\newcommand{\SP}{\mbox{ SPAN}}
\newcommand{\GF}{\mbox{ GF}}
\newcommand{\gf}{\mbox{ {\tiny{GF}}}}
\newcommand{\GR}{\mbox{ GR}}
\newcommand{\Di}{\mbox{ DIV}}
\newcommand{\mt}{\hspace{10mm} \mbox{ }}
\newcommand{\mf}{\hspace{5mm} \mbox{ }}
\newcommand{\mz}{\hspace{2mm} \mbox{ }}
\newcommand{\ga}{\gamma}
\newcommand{\xl}{\begin{tiny} \begin{array}{c} < \\ \simeq \end{array} \end{tiny}}
\newcommand{\xm}{\begin{tiny} \begin{array}{c} > \\ \simeq \end{array} \end{tiny}}
\newcommand{\OR}{\mbox{ ord}}
\newcommand{\dg}{\mbox{ deg}}
\newcommand{\QED}{\rule{2mm}{3mm}}
\newtheorem{thm}{Theorem}
\newtheorem{lem}{Lemma}
\newtheorem{cj}{Conjecture}
\newtheorem{cor}{Corollary}
\newtheorem{df}{Definition}
\newtheorem{imp}{Implication}
\newcommand{\mo}{\mbox{ mod }}
\newcommand{\Tr}{\mbox{ Tr}}
\newcommand{\mn}{\mbox{ {\tiny{min}}}}
\newcommand{\mx}{\mbox{ {\tiny{max}}}}
\newcommand{\erf}{\mbox{ erf}}
\newcommand{\erfc}{\mbox{ erfc}}
\newcommand{\SNR}{\mbox{ SNR}}
\newcommand{\BER}{\mbox{ BER}}
\newcommand{\hl}{\\ \hline}

\begin{tiny}
\end{tiny}

\cb

{\large{\bf Quantum Factor Graphs}}

{\sl Matthew.G.Parker \begin{footnote}{This work was funded by NFR Project Number 119390/431, and 
was presented in part at
2nd Int. Symp. on Turbo Codes and Related Topics, Brest, Sept 4-7, 2000}\end{footnote}
\\
     Inst. for Informatikk,
         University of Bergen, \\
          5020 Bergen, Norway, }

          {\tt matthew@ii.uib.no
          \\ http://www.ii.uib.no/$\sim$matthew/MattWeb.html
          }
\ce

{\bf{Biography:}} Matthew Parker is currently a Postdoctoral Researcher in the Code Theory Group at the
University of Bergen, Norway. Prior to this he was a Postdoctoral Researcher in the Telecommunications
Research Group at the University of Bradford, UK. His research interests are Iterative Computation, Sequence Design,
Quantum Computation, and Coding Theory.

\newpage

{\bf{Abstract: }} {\em The natural Hilbert Space of quantum particles
can implement maximum-likelihood (ML) decoding of classical information.
The 'Quantum Product Algorithm' (QPA) is computed on a Factor Graph,
where function nodes are unitary matrix operations followed by appropriate
quantum measurement. QPA is like the Sum-Product Algorithm (SPA), but without summary, giving
optimal decode with exponentially finer detail than achievable using SPA.
Graph cycles have no effect on QPA performance. QPA must be repeated a number of
times before successful and the ML codeword is obtained only after
repeated quantum 'experiments'. ML amplification improves decoding accuracy, and Distributed QPA facilitates
successful evolution.}

\vspace{5mm}


\newpage

\section{Introduction}
\label{sec1}
Recent interest in Turbo Codes \cite{Berr:Tur} and Low Density Parity Check Codes \cite{Gall:LDPC,Mac:LDPC}
has fuelled development of Factor Graphs and
associated Sum-Product Algorithm \cite{Ksc:Fac,McE:GDL1} (SPA), with applications to error-correction,
signal processing, statistics, neural networks,
and system theory.
Meanwhile the possibility of
Quantum Computing has sparked much interest \cite{Sho:Qua,Stea:QC}, and Quantum Bayesian Nets have
been proposed to help analyse and design Quantum Computers \cite{Tuc:QB,Tuc:QB2}.
This paper links these areas of research, showing that quantum resources
can achieve maximum-likelihood (ML) decoding of classical information. The natural Hilbert Space of a quantum
particle encodes a probability vector, and the joint-state of quantum particles
realises the 'products' associated with SPA. SPA summary is omitted as quantum bits (qubits) naturally
encode the total joint-probability state.
Dependencies between vector indices become 'entanglement' in quantum space, with the
Factor Graph defining dependency (entanglement) between qubits. Graph function nodes are implemented
as unitary matrix
\begin{footnote}{
'Unitary' means that ${\bf{U}}$ satisfies $ {\bf{UU^{\dag}}} = {\bf{I}} $,
where $\dag$ means 'conjugate transpose'.}
\end{footnote}
-vector products followed by quantum measurement. This is the Quantum Product
Algorithm (QPA). As QPA avoids summary it
 avoids problems encountered by SPA on graphs with short cycles. Moreover, whereas SPA is
 iterative, using message-passing and activating each node more than once, QPA does not
iterate but must successfully activate each node only once. However the (severe) drawbacks with QPA are
 as follows: 1) Each function node must be repeatedly activated until it successfully
'prepares' it's local variable nodes (qubits) in the correct entangled state - any activation
failure destroys evolution in all variable nodes already entangled with local
variables.
2) Once a complete Factor Graph has successfully evolved, final quantum measurement
only delivers the ML codeword with a certain (largest) probability. Repeated successful evolutions
then determine the ML codeword to within any degree of confidence.
This second drawback can be overcome by suitable "ML Amplification" of QPA output, prior
to measurement.

Section \ref{sec2} presents QPA, highlighting its ability to deliver the optimal output
joint-state, unlike SPA.
Quantum systems describe the exact joint-state by appropriate
'entanglement' with and measurement of ancillary qubits.
Section \ref{sec3} considers a simple example of QPA on Quantum Factor Graphs, showing
that iteration on graphs with cycles is unnecessary because QPA avoids premature
summary.
Section \ref{sec4} shows how to amplify the likelihood of measuring the ML codeword from
QPA output.
Unfortunately QPA must be repeated many times and/or executed in parallel to have a
hope of successful completion. Suitable distributed QPA scheduling is discussed in Section \ref{sec5},
and it is argued that successful QPA completion is conceivable using asynchronous distributed processing on
many-node Factor Graphs. This paper does not deal with phase properties of quantum computers. It is expected that
the inclusion of phase and non-diagonal unitary matrices will greatly increase functionality of the Quantum
Factor Graph. 

The aim of this paper is {\underline{not}} to propose an immediately realisable implementation of a quantum
computer. Rather, it is to highlight similarities between graphs for classical message-passing, and
graphs that 'factor' quantum computation. The paper also highlights the differences between the two graphs:
whereas classical graphs can only ever compute over a tensor product space, the quantum graph can
compute over the complete entangled (tensor-irreducible) space.
\section{The Quantum Product Algorithm (QPA)}
\label{sec2}
\subsection{Preliminaries}
\begin{figure}[h]
\centerline{
\epsfig{file=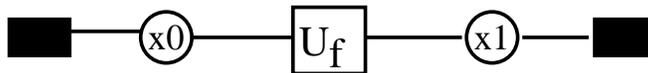,height=1cm}
}
\caption{Two-Qubit Factor Graph}
\label{fg1}
\end{figure}

Consider the Factor Graph of Fig \ref{fg1}.

Let ${\bf{U_f}} = \left (
\begin{array}{cccc}
f_0 & 0 & 0 & 0 \\
0   & f_1 & 0 & 0 \\
0   & 0   & f_2 & 0 \\
0   & 0   & 0   & f_3
\end{array}
\right )$, and
${\bf{U_g}} = \left (
\begin{array}{cccc}
g_0 & 0 & 0 & 0 \\
0   & g_1 & 0 & 0 \\
0   & 0   & g_2 & 0 \\
0   & 0   & 0   & g_3
\end{array}
\right )
\begin{array}{c}
\m{, where $|g_k|^2 = 1 - |f_k|^2$,} \\ \m{and $f_k^*g_k + f_kg_k^* = 0$, $\forall k$.} \\
\m{'$*$' means complex conjugate.}
\end{array}$ Let
${\bf{U_{fg}}} = \left (
\begin{array}{cc}
{\bf{U_f}} & {\bf{U_g}}  \\
{\bf{U_g}}   & {\bf{U_f}}
\end{array}
\right )$. ${\bf{U_{fg}}}$ is unitary, and the ${\bf{U_f}}$ of Fig \ref{fg1} and subsequent figures
always implies the action of ${\bf{U_{fg}}}$ together with the measurement of an ancillary qubit, $z$,
as described below.
A qubit, $x_i$, can be in states $0$ or $1$ or in a
statistical superposition of $0$ and $1$.
Let qubits $x_0,x_1$ be initialised (by the black boxes) to states $x_0 = (\al_0,\beta_0)^T$ and
$x_1 = (\al_1,\beta_1)^T$, where $\al_i,\beta_i$ are complex probabilities such that
$|\al_i|^2 + |\beta_i|^2 = 1$. For instance, $x_0$ is in states $0$ and $1$ with probabilities
$|\al_0|^2$ and $|\beta_0|^2$, respectively. Let an ancillary qubit, $z$, be initialised
to state $0$, i.e. $z = (1,0)$. Then the initial joint probability product-state of
qubits $x_0,x_1,z$ is ${\bf{A}} = (\al_0,\beta_0)^T \otimes (\al_1,\beta_1)^T \otimes (1,0)^T =
(\al_0\al_1,\beta_0\al_1,\al_0\beta_1,\beta_0\beta_1,0,0,0,0)^T =
(s_0,s_1,s_2,s_3,0,0,0,0)^T$, where $|s_0|^2 + |s_1|^2 + |s_2|^2 + |s_3|^2 = 1$, and '$\otimes$' is the
tensor product.
The element at vector index $v$ is the probability that the qubits are in state
$v$. For instance, qubits
$x_0x_1z$ are in joint-state $010$ with probability $|s_2|^2$.
Subsequent measurement of a subset of the
qubits
projects the measured qubits to a fixed substate with a certain probability, and 'summarises' the
vector for the remaining non-measured qubits.
Thus QPA is as follows,
\bite
    \item Compute ${\bf{S}} = {\bf{U_{fg}A}}$.
    \item Measure qubit $z$. With probability
    $p_f = |s_0f_0|^2 + |s_1f_1|^2 + |s_2f_2|^2 + |s_3f_3|^2$
        we collapse $z$ to $0$, and $x_0,x_1$ to joint-state
        ${\bf{S_f}} = \mu_0
(s_0f_0,s_1f_1,s_2f_2,s_3f_3)^T$. With probability
    $p_g = |s_0g_0|^2 + |s_1g_1|^2 + |s_2g_2|^2 + |s_3g_3|^2$
        we collapse $z$ to $1$, and $x_0,x_1$ to joint-state
        $ {\bf{S_g}} = \mu_1(s_0g_0,s_1g_1,s_2g_2,s_3g_3)^T$. $\mu_0$ and $\mu_1$ are normalisation
        constants. $p_f + p_g = 1$. ${\bf{S_f}}$ is our desired QPA result.
Successful
        QPA completion is self-verified when we measure $z = 0$.
\eite
In contrast, classical SPA computes ${\bf{S_f}} = {\bf{U_fA}}$ (with probability 1) and
must then perform a subsequent 'summary' step on ${\bf{S_f}}$ before returning a result
for each variable separately. This result is, \newline
$x_0 = |\mu_0|^2(|s_0f_0|^2 + |s_2f_2|^2,|s_1f_1|^2 + |s_3f_3|^2)^T$,
$x_1 = |\mu_1|^2(|s_0f_0|^2 + |s_1f_1|^2,|s_2f_2|^2 + |s_3f_3|^2)^T$. \newline
For instance, for $x_0 = 0$ we sum the two {\bf{classical}}
\begin{footnote}
{Classical SPA probabilities in this paper are always represented as the magnitude-squared of
their quantum counterparts}
\end{footnote}
probabilities of ${\bf{S_f}}$ where $x_0 = 0$ to get $|s_0f_0|^2 + |s_2f_2|^2$. Similarly, for
$x_0 = 1$ we summarise to $|s_1f_1|^2 + |s_3f_3|^2$. It is in this sense that SPA
is a 'tensor-approximation' of QPA.

We identify the following successively accurate computational scenarios (decoding modes)
for a space of $N$ binary-state variables:
\bite
	\item {\bf{Hard-Decision}} operates on a probability space, \newline
$(\al_0,\beta_0) \otimes (\al_1,\beta_1) \otimes \ldots \otimes (\al_{N-1},\beta_{N-1}), \mf \al,\beta \in \{0,1\}$
        \item {\bf{Soft-Decision}} operates on a probability space, \newline
$(\al_0,\beta_0) \otimes (\al_1,\beta_1) \otimes \ldots \otimes (\al_{N-1},\beta_{N-1}), \mf \al,\beta \in \{\m{Real Numbers $0 \rightarrow 1$}\}$
        \item {\bf{Quantum Soft-Decision}} operates on a probability space, \newline
$(\al_0,\beta_0) \otimes (\al_1,\beta_1) \otimes \ldots \otimes (\al_{N-1},\beta_{N-1}), \mf \al,\beta \in \{\m{Complex Numbers}\}$
        \item {\bf{Entangled-Decision}} operates on a probability space, \newline
$(s_0,s_1,s_2,\ldots,s_{2^N-1})$, $s \in \{\m{Complex Numbers}\}$
\eite
All four of the above Decision modes satisfies the probability restriction that the sum of the magnitude-squareds
of the vector elements is 1.
Both Quantum Soft-Decision and Entangled-Decision make use of the natural quantum statistical properties of
matter, including the property of Superposition. Moreover, Entangled-Decision operates over exponentially
larger space. Classical SPA operates in Soft-Decision mode. QPA operates in Entangled-Decision mode.
In the previous discussion it was assumed that the QPA was operating on input of the form,
$(\al_0,\beta_0)^T \otimes (\al_1,\beta_1)^T \otimes (1,0)^T$. More generally, QPA can operate on input and deliver
output in Entangled-Decision mode. This is in strong contrast to SPA which must summarise both input and output
down to Soft-Decision mode. It is this approximation that forces SPA to iterate and to sometimes fail on graphs
with cycles.

Consider the following example. If the diagonal of ${\bf{U_f}}$ is $(1,0,0,1)$, then ${\bf{U_f}}$ represents
XOR, and Fig \ref{fg1} decodes to codeset ${\bf{C}} = \{00,11\}$ (i.e. $x_0 + x_1 = 0, \mo 2$).
${\bf{C}}$ has distance 2, which is optimal for length 2 binary codes: in general
if ${\bf{U_f}}$ cannot be tensor-decomposed then it represents a code ${\bf{C}}$
with good distance properties.
Initially,
let $x_0 = (\sqrt{0.4},\sqrt{0.6})^T$, $x_1 = (\sqrt{0.6},\sqrt{0.4})^T$. Then
${\bf{A}} = (\sqrt{.24},0.6,0.4,\sqrt{0.24},0,0,0,0)^T$, and
${\bf{S_f}} = \frac{1}{\sqrt{2}}(1,0,0,1)^T$. $p_f = 0.48$, so, on average, $48$
${\bf{S_f}}$ outputs are computed for every $100$ QPA attempts. The ML codeword
is both $00$ and $11$, and when ${\bf{S_f}}$ is measured, $00$ and $11$ are equally likely to
be returned. In contrast, classical SPA for the same input returns
$x_0 = x_1 = (\frac{1}{2},\frac{1}{2})$,
implying (wrongly) an equally likely decode to any
of the words $00,01,10,11$. So even in this simplest example the advantage of QPA over SPA is evident.
\subsection{Product Space for Classical SPA}
Because $x_0$ and $x_1$ are separated in Fig \ref{fg1}, their classical
joint-state only represents tensor {\bf{product}} states (Soft-Decision mode). An equivalent Factor Graph to
that of Fig
\ref{fg1}
could combine $x_0$ and $x_1$ into one quaternary variable which would reach all non-product
quaternary states. But this requires 'thickening' of graph communication lines and
exponential increase in SPA computational complexity. Consequently only limited variable 'clustering'
is desirable, although too little
clustering 'thins out' the solution space to an insufficient highly-factored product space. This is the
fundamental Factor Graph
trade-off - good Factor Graphs achieve efficient SPA by careful variable 'separation', ensuring the
joint product space is close enough to the exact (non-summarised) non-product space.

\subsection{Entangled Space for QPA}
In contrast, although $x_0$ and $x_1$ are physically separated in Fig \ref{fg1}, quantum non-locality
must take into account correlations between $x_0$ and $x_1$. Their joint-state
now occurs over the union of product and (much larger) non-product (entangled) space (Entangled-Decision mode).
An entangled joint-state
vector cannot be tensor-factorised over constituent qubits. QPA does not
usually output to product space because the joint-state of output qubits is usually entangled. In fact
QPA is algorithmically simpler than SPA, as SPA is a subsequent tensor approximation
of QPA output at each local function.
\subsection{Example}
Let the diagonal of ${\bf{U_f}}$ be $(1,0,0,1)$.
Initialise $x_0$ and $x_1$ to
joint-product-state,
$x_0 = \frac{1}{\sqrt{3}}(1,\sqrt{2})^T$, $x_1 = \frac{1}{\sqrt{2}}(1,1)^T$.
With probability $p_f = 0.5$ QPA measures $z = 0$ and computes the
joint-state of $x_0,x_1$ as
$ {\bf{S_f}} = \frac{1}{\sqrt{3}}
(1,0,0,\sqrt{2})^T$.
A final measurement of qubits $x_0$ and $x_1$ yields codewords $11$
and $00$
with probability $\frac{2}{3}$, and $\frac{1}{3}$, respectively. In contrast
SPA summarises ${\bf{S_f}}$ to
$x_0 = x_1 = \frac{1}{3}(1,2)$. Although
a final 'hard-decision' on $x_0$ and $x_1$ chooses, correctly, the ML codeword $x_0 = x_1 = 1$,
the joint-product-state output, $\frac{1}{3}(1,2)^T \otimes
\frac{1}{3}(1,2)^T =
\frac{1}{9}(1,2,2,4)^T$ assigns, incorrectly, a non-zero probability to words $01$ and $10$.
\subsection{A Priori Initialisation}
To initialise $x_0$ to $(\al_0,\beta_0)^T$, we again use QPA. Let the diagonal of
${\bf{U_f}}$ (for the left-hand black box of Fig \ref{fg1}) be $(\al_0,\beta_0)$. Then
the diagonal of ${\bf{U_g}}$ is $\pm i(\frac{\al_0 \sqrt{ 1 - |\al_0|^2}}{|\al_0|},
\frac{\beta_0 \sqrt{ 1 - |\beta_0|^2}}{|\beta_0|})^T$. Measurement of $z = 0$
initialises $x_0$ to $(\al_0,\beta_0)^T$, and this occurs with probability $p_f = 0.5$.
$x_1$ is initialised likewise.

\subsection{Comments}
The major drawback of QPA is the significant probability of QPA failure,
occurring when $z$ is measured as $1$. This problem is amplified for larger Quantum
Factor Graphs where a different $z$ is measured at each local function;
{\bf{QPA evolution failure at a function node not only destroys the states of
variables connected with that function, but also destroys all states of variables entangled
with those variables.}} QPA is more likely to succeed when input variable probabilities
are already skewed somewhat towards a valid codeword.
Section \ref{sec3} shows how QPA can operate successfully
even when SPA fails.
\section{Quantum Product Algorithm on Factor Graphs with Cycles}
\label{sec3}
\begin{figure}[h]
\centerline{
\epsfig{file=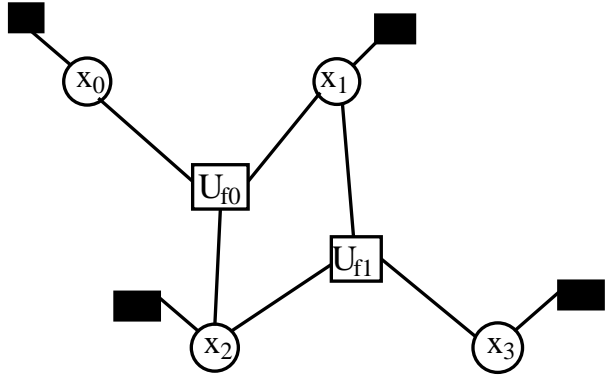,height=5cm}
}
\caption{Factor Graph with a Cycle}
\label{fg2}
\end{figure}

This section shows that graph cycles do not compromise QPA performance.
Consider the Factor Graph of Fig \ref{fg2}.

Functions ${\bf{U_{f0}}}$ and ${\bf{U_{f1}}}$ are both $8 \times 8$ XOR diagonal matrices with diagonal elements
$(10010110)$. Acting on the combined four-qubit space, $x_0x_1x_2x_3$, they are the functions
${\bf{U_{f0}}} \otimes {\bf{I_2}}$ and
${\bf{I_2}} \otimes {\bf{U_{f1}}}$, respectively, with diagonal
elements $(1001011010010110)$ and $(1100001100111100)$, respectively, where ${\bf{I_2}}$ is the
$2 \times 2$ identity matrix. QPA on
Fig \ref{fg2} performs the global function
${\bf{U_F}} = \left ( {\bf{U_{f0}}} \otimes {\bf{I_2}} \right ) \left ( {\bf{I_2}} \otimes {\bf{U_{f1}}} \right )$
on four-qubit space, with diagonal elements $(1000001000010100)$, forcing
output into codeset ${\bf{C}} = \{ 0000,0110,1011,1101 \}$. Functions ${\bf{U_{f0}}}$, ${\bf{U_{f1}}}$,
and ${\bf{U_F}}$ 'sieve' the input joint-state, where ${\bf{U_F}}$ is
the combination of two 'sub-sieves', ${\bf{U_{f0}}}$ and ${\bf{U_{f1}}}$. QPA iteration
(i.e. successfully completing a sub-function
more than once on the same qubits) has no purpose, as only one needs apply a particular sieve once. So
graph cycles have no bearing on QPA. (However iteration may be useful to maintain the entangled
result in the presence of quantum decoherence and noise).
To underline cycle-independence, consider the action of SPA, then QPA on Fig \ref{fg2}.

Initialise as follows (using classical probabilities),
$$ x_0x_1x_2x_3 = (0.1,0.9)^T \otimes (0.6,0.4)^T \otimes (0.6,0.4)^T \otimes (0.6,0.4)^T $$
Hard-decision gives $x_0x_1x_2x_3 = 1000$, which can then be decoded
algebraically to codeword $0000$. However optimal soft-decision would
decode to either $x_3x_2x_1x_0 = 1011$ or $1101$, with equal probability. Because of the
small graph cycle SPA fails to decode correctly, and settles to the
joint-product-state, \newline
$x_0x_1x_2x_3 =
(0.108,0.892)^T \otimes (0.521,0.479)^T \otimes (0.521,0.479)^T \otimes (0.601,0.399)^T$.
A final hard-decision on this
output gives non-codeword $x_0x_1x_2x_3 = 1000$ which can then be decoded
algebraically, again to codeword $0000$.
In contrast, successful QPA outputs the optimal entangled joint-state,
\newline
$S_F = \frac{1}{\sqrt{2040}}(\sqrt{216},0,0,0,0,0,\sqrt{96},0,0,0,0,\sqrt{864},0,\sqrt{864},0,0)^T$.
Final measurement of $S_F$ always outputs a codeword from ${\bf{C}}$, and with
probability $\frac{2 * 864}{2040}$ outputs either $1011$ or $1101$.
QPA evolves on Fig \ref{fg2} correctly with probability $0.204$. Therefore $1000$
attempts produce around $204$ correctly entangled joint-states.

\begin{figure}[h]
\centerline{
\epsfig{file=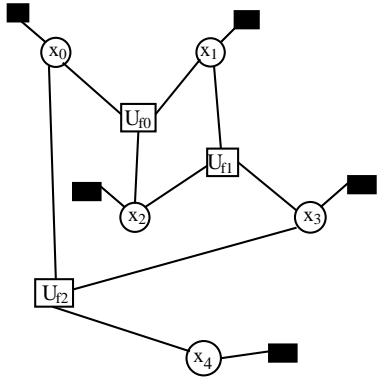,height=5cm}
}
\caption{Extended Factor Graph with a Cycle}
\label{fg3}
\end{figure}

To underline QPA advantage, consider the single variable extension of Fig \ref{fg2}
in Fig \ref{fg3}, where $x_4$ is initialised to $(\sqrt{0.5},\sqrt{0.5})^T$.

As $x_4 = x_0 \oplus x_3$, and our original code, ${\bf{C}} = \{ 0000,0110,1011,1101
\}$, always had $x_0 = x_3$, then $x_4$ should always be $0$. But SPA on
Fig \ref{fg3} computes $x_4 = (0.421,0.579)^T$ and subsequent hard-decision gives
$x_4 = 1$. In contrast, successful QPA computes the optimal non-product joint-state,

\begin{tiny}
$$ S_{F'} = \frac{1}{\sqrt{2040}}(\sqrt{216},0,0,0,0,0,\sqrt{96},0,0,0,0,\sqrt{864},0,\sqrt{864},0,0,
0,0,0,0,0,0,0,0,0,0,0,0,0,0,0,0)^T$$
\end{tiny}

Final measurement of $S_{F'}$ {\bf{always}} outputs $x_4 = 0$.
QPA evolves on Fig \ref{fg3} correctly with probability $0.204 * 0.5 = 0.114$.
\section{Maximum-Likelihood (ML) Amplification}
\label{sec4}
\subsection{Preliminaries}
The ML codeword is the one most likely to be measured from QPA output, with
probability, $p_M$, say. For instance, if QPA output of Fig \ref{fg1} is
$S_f = \frac{1}{\sqrt 3}(1,0,0,\sqrt 2)^T$, say, then $11$ is the ML codeword, and it is
measured with probability $p_M = \frac{2}{3}$. Numerous executions of QPA on the same input
will verify that $11$ is, indeed, the ML codeword. However these numerous executions must output to a
length $2^N$ final averaging probability vector (for $N$ qubits). We do not want to
store such an exponential vector. Instead, therefore, we
'amplify' the statistical advantage of $11$ over $00$ prior to measurement,
thereby making $11$ significantly more likely to be read. This is achieved
by computing the square of each quantum vector element as follows.
Consider two independent
QPA executions on the same input, both outputting $S_f$.
Associate these outputs with qubits $x_{0,0},x_{1,0}$, and $x_{0,1},x_{1,1}$.
The joint-state of qubits $x_{0,0},x_{1,0},x_{0,1},x_{1,1}$ is,
$$ V_0 = S_f \otimes S_f =
\frac{1}{3}(1,0,0,\sqrt 2,0,0,0,0,0,0,0,0,\sqrt 2,0,0,2)^T $$
Consider the unitary permutation matrix
\begin{tiny}
$$ {\bf{P}} = \left ( \begin{array}{c}
1000000000000000 \\
0000010000000000 \\
0000000000100000 \\
0000000000000001 \\
0100000000000000 \\
0010000000000000 \\
0001000000000000 \\
0000100000000000 \\
0000001000000000 \\
0000000100000000 \\
0000000010000000 \\
0000000001000000 \\
0000000000010000 \\
0000000000001000 \\
0000000000000100 \\
0000000000000010
\end{array} \right ) $$
\end{tiny}
Only the '1' positions in the first four rows are important. Performing
${\bf{P}}$ on $x_{0,0},x_{1,0},x_{0,1},x_{1,1}$, gives,
$$ {\bf{P}}V_0 = \frac{1}{3}(1,0,0,2,0,0,\sqrt 2,0,0,0,0,0,0,\sqrt 2,0,0)^T $$
We then measure qubits $x_{0,1},x_{1,1}$. With probability $p_{a_0} = \frac{5}{9}$
we read $x_{0,1} = x_{1,1} = 0$, in which case $x_{0,0}$ and $x_{1,0}$ are forced
into joint state $S_{f,1} = \frac{1}{\sqrt 5}(1,0,0,2)$, which is the element-square of
$S_f$. A measurement of $S_{f,1}$ returns $11$ with probability $p_M = \frac{4}{5}$, which
is a significant improvement over $p_M = \frac{2}{3}$. Likewise we compute the
element fourth-powers of $S_f$ by preparing two independent qubit pairs in $S_{f,1}$ and
permuting the (umeasured) joint state vector $V_1 = S_{f,1} \otimes S_{f,1}$ to give
$ {\bf{P}}V_1$, and then measuring the second pair of qubits. With
probability $p_{a_1} = \frac{17}{25}$ we read this pair as $00$, in which case the
first two qubits are forced into the joint-state $S_{f,2} = \frac{1}{\sqrt
17}(1,0,0,4)$, which is the element fourth-power of
$S_f$. A measurement of $S_{f,2}$ returns $11$ with probability $p_M = \frac{16}{17}$, which
is a further improvement over $p_M = \frac{2}{3}$. In this way we amplify the
likelihood of measuring the ML codeword. To compute the element $2^k$\thh-power, $S_{f,k}$, we
require, on average, $\frac{2}{p_{a_k}}$ independent preparations, $S_{f_{k-1}}$, each of
which requires, on average, $\frac{2}{p_{a_{k-1}}}$ independent preparations,
$S_{f_{k-2}}$, and so on.

We can perform QPA on large Factor Graphs, then amplify the result $k$
times to ensure a high likelihood of measuring the ML codeword, as described above.
However the above amplification acts on the complete graph with one operation, ${\bf{P}}$.
It would be preferable to decompose ${\bf{P}}$ into $4 \times 4$ unitary matrices which
only act on independent qubit pairs $x_{i,0}$ and $x_{i,1}$, thereby localising
amplification. Consider, once again, Fig \ref{fg1}. From the point of view of
$x_{0,1}$, $x_{0,0}$ appears to be in summarised state
\begin{footnote}{$x_{0,0}$ is generally {\underline{not}} in this summarised state,
due to phase considerations, but the viewpoint is valid for our purposes as long as
subsequent unitary matrix operations on $x0$ only have one non-zero entry per row.}
\end{footnote},
$s_f = \frac{1}{\sqrt 3}(1,\sqrt 2)^T$. Similarly, from the point of view of
$x_{0,0}$, $x_{0,1}$ appears to be in state $s_f$. Thus $x_{0,0},x_{0,1}$ appear to be in
joint product state $v_0 = \frac{1}{3}(1,\sqrt 2,\sqrt 2,2)^T$. Consider unitary
permutation matrix,

\begin{tiny}
$$ {\bf{Q}} = \left ( \begin{array}{c}
1000 \\
0001 \\
0100 \\
0010
\end{array} \right ) $$
\end{tiny}
We compute
$ {\bf{Q}}v_0 = \frac{1}{3}(1,2,\sqrt 2,\sqrt 2)^T $ on qubits $x_{0,0},x_{0,1}$
and measure qubit $x_{0,1}$. With probability $p_{a_0} = \frac{5}{9}$
we read $x_{0,1} = 0$, in which case $x_{0,0}$ is forced
into joint state $s_{f,1} = \frac{1}{\sqrt 5}(1,2)$, which is the element-square of
$s_f$. Due to the exact form of our joint-state vector, $S_f$, this single measurement
is enough to also force $x_{0,0}x_{1,0}$ into joint state $S_{f,1}$. However, for a
general function $S_f$, we should perform ${\bf{Q}}$ on every qubit pair,
$x_{i,0}x_{i,1}$, then measure $x_{i,1}$ $\forall i$. This is equivalent to performing
${\bf{P'}} = {\bf{Q}} \otimes {\bf{Q}}$ on (re-ordered) joint-state
vector $x_{0,0}x_{0,1}x_{1,0}x_{1,1}$, and this is identical to performing
${\bf{P}}$ on $x_{0,0}x_{1,0}x_{0,1}x_{1,1}$. The probability of measuring
$x_{1,0} = x_{1,1} = 0$ is the same whether ${\bf{P}}$ or ${\bf{Q}}$ is used.
The same process is followed
to achieve element $2^k$th powers.
\subsection{The Price of Amplification}
\begin{figure}[h]
\centerline{
\epsfig{file=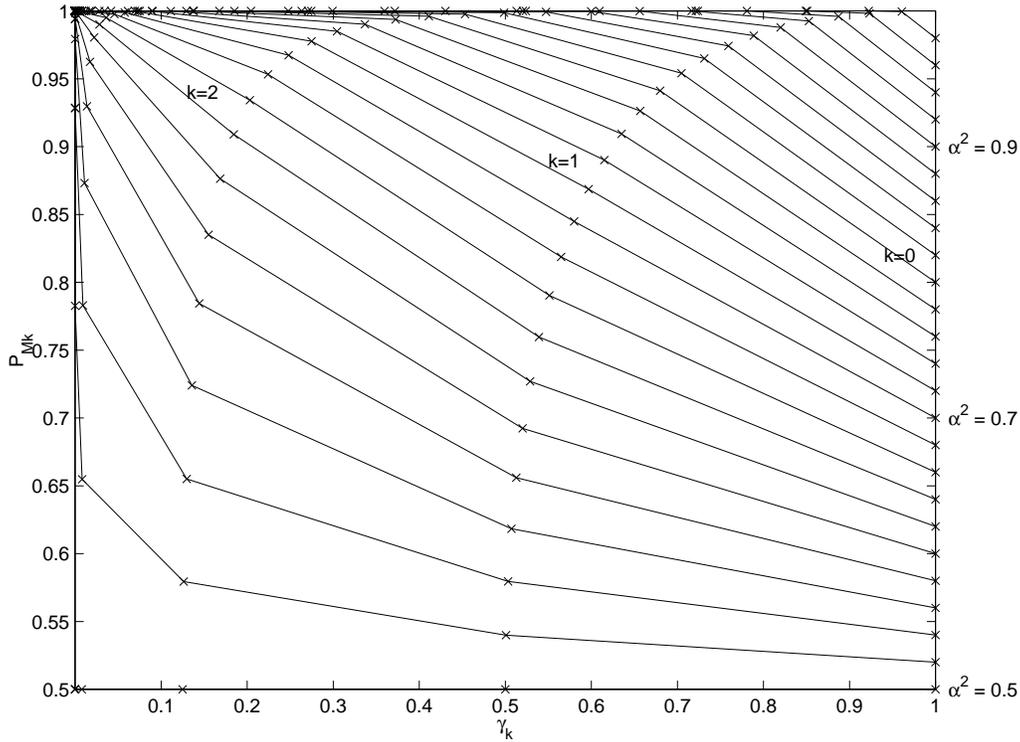,height=10cm}
}
\caption{Amplification Success Probability, $\gamma_k$, v ML Advantage, $P_{Mk}$}
\label{fg4}
\end{figure}

There is a statistical cost to qubit amplification.
Let $s = (\al,\beta)^T$ be the initial state of a qubit $x$, where, for notational convenience,
we assume that $\al$ and $\beta$ are both real. Then $\al^2 + \beta^2 = 1$ and, given $2^k$ qubits
all identically prepared in state $s$, the likelihood of preparing one qubit in (unnormalised) state
$s_k = (\al^{2^k},\beta^{2^k})^T$ is $\gamma_k$, where,
$$ \gamma_k = \gamma_{k-1}^2 \frac{r_{k+1}}{r_k^2}, \mf \gamma_0 = 1 $$
and $r_k = \al^{2^k} + \beta^{2^k}$. For a qubit in state $s_k$, the probability
of selecting the ML codebit is,
$$ P_{Mk} = \frac{\al^{2^{k+1}}}{\al^{2^{k+1}} + \beta^{2^{k+1}}} $$
(assuming $\al \ge \beta$). We can plot $\gamma_k$ against $P_{Mk}$ for various
$\al^2$ as $k$ varies, as shown in Fig \ref{fg4}.

Each of the 25 lines in Fig \ref{fg4} refers to a different value of $\al^2$, for $\al^2$ from
$0.5$ up to $0.98$ in steps of $0.2$. The initial state, $s$, when $k = 0$, occurs with probability
$\gamma_k = 1$, and is marked on the right-hand side of Fig \ref{fg4} for each of the 25 lines. After
one amplification step, $k = 1$, and another 25 points are marked on the graph to the left of the
points for $k = 0$, indicating that a successful amplification step has occurred with probability
$\gamma_k \le 1$. Similarly points for $k = 2$, $k = 3$,..etc are marked successively to the left on
Fig \ref{fg4}. The $y$-axis shows the ML advantage, $P_{Mk}$, which can be achieved with probability
$\gamma_k$ after $k$ steps for each value of $\al^2$.
For instance, when $s = (\al,\beta)^T = (\sqrt{0.62}, \sqrt{0.38})^T$,
then an ML advantage of $P_{Mk} = 0.9805$
can be ensured after $k = 3$ steps, and this can be achieved with probability $\gamma_k = 0.0223$
given $2^3 = 8$ independently prepared qubits, all in state $s$. Amplification is more rapid
if $s$ already has significant ML advantage (i.e. when $\al$ is high). In contrast if $\al^2 = 0.5$
then no amplification of
that qubit is possible. This is quite reasonable as, in this case, both states $0$ and $1$ are
equally likely, so there is no ML state. Successive measurement of zero of all second qubits
of each qubit pair self-verifies that we have obtained successful amplification. If, at any step, $k$,
the second qubit of the qubit pair
is measured as one then amplification fails and the graph local to this qubit which has been
successfully entangled up until now is completely destroyed.
\section{Distributed QPA on Many-Node Factor Graphs}
\label{sec5}
\subsection{Preliminaries}
In classical systems it is desirable to implement SPA on
Factor Graphs which 'tensor-approximate' the variable space using many small-state
variables (e.g. bits), linked by small-dimensional constituent functions, thereby minimising computational
complexity. In quantum systems  it is similarly desirable to implement QPA on
Factor Graphs using many small-state
variables (e.g. qubits), linked by small-dimensional constituent unitary functions. Any
Quantum Computation can be decomposed into a sequence of one or two-bit 'universal' gate unitary operations
\cite{DiV:Uni}\begin{footnote}{This also implies that any classical Factor Graph can be similarly
decomposed.}
\end{footnote}. Computational
complexity is minimised by using small-dimensional unitary matrices for constituent functions.
Moreover, fine granularity of the Factor Graph allows distributed node processing. This appears to be
essential for large Quantum Factor Graphs to have acceptable probability of successful global evolution, as
we will show.
Distributed QPA allows variable nodes to evolve entanglement only with neighbouring
variable nodes so that, if a local function measurement or amplification is unsuccessful, only local evolution
is destroyed. Remember that local evolution is OFTEN unsuccessful, as failure occurs when a local ancillary qubit,
$z$, is measured as $1$, or when a local amplifying qubit is measured as 1. Therefore node localities with high
likelihood of successful evolution (i.e. with positively skewed input probabilities) are likely to
evolve first. These will then encourage other self-contradictory node localities to evolve successfully. In contrast,
non-distributed QPA on large Factor Graphs using one large global function is very unlikely to ever succeed,
especially for
graphs encoding low-rate codes.

\begin{figure}[h]
\centerline{
\epsfig{file=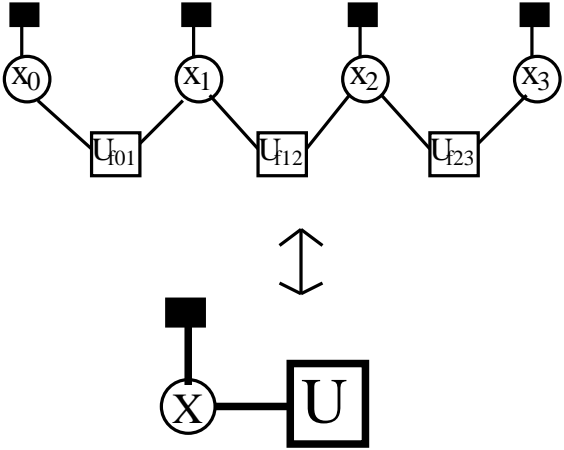,height=6cm}
}
\caption{Distributed QPA (top) Non-Distributed QPA (bottom), 4-bit code}
\label{fg5}
\end{figure}

To illustrate the advangtage of distributed QPA, consider the low rate code
of Fig \ref{fg5}, where ${\bf{U_{fij}}} = \m{diag}(1,0,0,1)$.
Both top and bottom graphs represent the code ${\bf{C}} = \{0000,1111\}$, where ${\bf{U}}$ is
a combination of XOR sub-matrices, ${\bf{U_{f01}}}$,${\bf{U_{f12}}}$, and ${\bf{U_{f23}}}$.
The top graph distributes processing.
We allow ${\bf{U_{f01}}}$ and ${\bf{U_{f23}}}$ to operate independently and in parallel.
Moreover, if ${\bf{U_{f01}}}$ fails to establish, then it does not destroy any successful evolution of
${\bf{U_{f23}}}$, as the two localities are not currently entangled. Once both
${\bf{U_{f01}}}$ and ${\bf{U_{f23}}}$ have completed successfully,
the subsequent probability of successful completion of ${\bf{U_{f12}}}$ is, in general, likely to increase.
So distributing QPA increases likelihood of successful evolution of the complete Factor Graph.
We now demonstrate this graphically. Let qubits $x_0,x_1,x_2,x_3$ of Fig \ref{fg5} initially be in
states $x_0 = (\al_0,\beta_0)^T$, $x_1 = (\al_1,\beta_1)^T$, $x_2 = (\al_2,\beta_2)^T$,
$x_3 = (\al_3,\beta_3)^T$, where, for notational convenience, we assume all values are real. Then
$\al_i^2 + \beta_i^2 = 1$, $\forall i$.
The probability of successful completion of ${\bf{U_{f01}}}$ is
$p_{f01} = (\al_0\al_1)^2 + (\beta_0\beta_1)^2$, and probability of successful completion of
${\bf{U_{f23}}}$ is $p_{f23} = (\al_2\al_3)^2 + (\beta_2\beta_3)^2$.
Therefore the probability of successful completion of both ${\bf{U_{f01}}}$ and ${\bf{U_{f23}}}$
after exactly $q$ parallel attempts (no less) is,

\begin{small}
$$ p_{0-3}(q) = (1 - p_{f01})^{q-1}(1 - p_{f23})^{q-1}p_{f01}p_{f23} +
(1 - p_{f01})^{q-1}(1 - (1 - p_{f23})^{q-1})p_{f01} + (1 - p_{f23})^{q-1}(1 - (1 - p_{f01})^{q-1})p_{f23} $$
\end{small}

\begin{figure}[h]
\centerline{
\epsfig{file=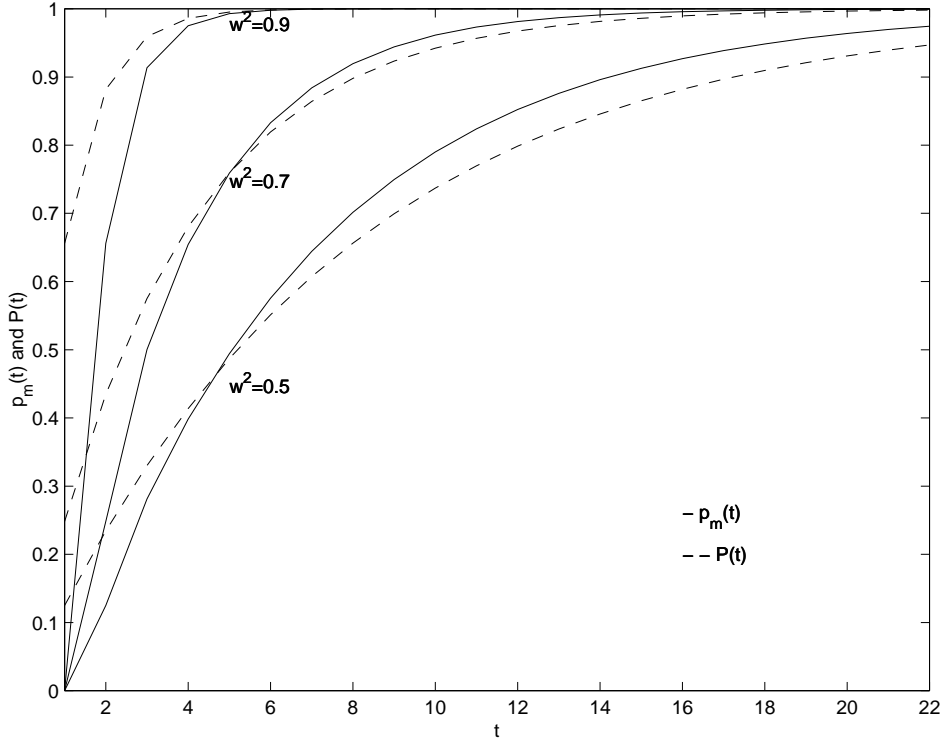,height=10cm}
}
\caption{No of Steps v Non-Distributed and Distributed QPA: Completion Probabilities}
\label{fg6}
\end{figure}

\begin{figure}[h]
\centerline{
\epsfig{file=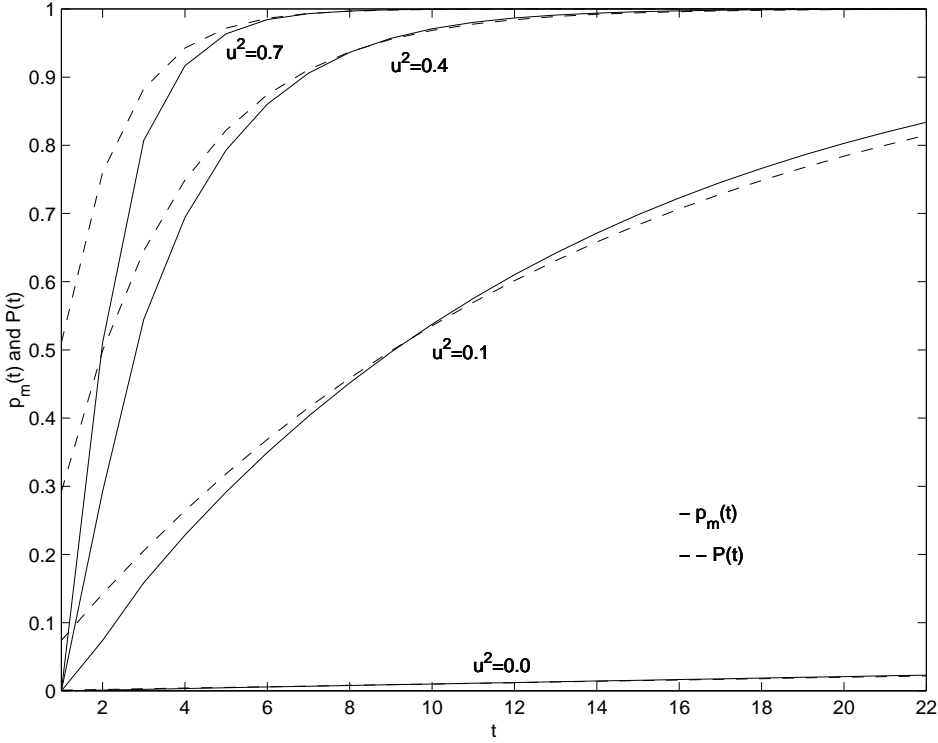,height=10cm}
}
\caption{No of Steps v Non-Distributed and Distributed QPA: $w^2 = 0.9$, $\al_0$ varies}
\label{fg7}
\end{figure}

Given successful completion of ${\bf{U_{f01}}}$ and ${\bf{U_{f23}}}$,
the probability of {\bf{subsequent}} successful completion of ${\bf{U_{f12}}}$ is,
$$ p_{f12}' = \frac{(\al_0\al_1\al_2\al_3)^2 + (\beta_0\beta_1\beta_2\beta_3)^2}{p_{f01}p_{f23}} $$
Therefore the probability of successful completion of ${\bf{U_{f01}}}$ and ${\bf{U_{f23}}}$,
immediately followed by successful completion of ${\bf{U_{f12}}}$ is,
$ p_{0\rightarrow 3}(q) = p_{0-3}(q-1)p_{f12}'$, and the probability of successful completion of ${\bf{U_{f01}}}$ and
${\bf{U_{f23}}}$, immediately followed by completion failure of ${\bf{U_{f12}}}$ is,
$ {\overline{p_{0\rightarrow 3}}}(q) = p_{0-3}(q-1)(1 - p_{f12}')$.
Therefore the probability of successful completion after exactly $t$ steps of ${\bf{U_{f01}}}$ and
${\bf{U_{f23}}}$ in parallel, followed by ${\bf{U_{f12}}}$, is,
$$ p_e(t) = \sum_{q=2}^t p_{0\rightarrow 3}(q) \sum_{{\bf{v \in D(t-q)}}} \prod_{u \in {\bf{v}}}
{\overline{p_{0\rightarrow 3}}}(u) $$ where ${\bf{D(k)}}$ is the set of unordered partitions of
$k$. Therefore the probability of successful completion after at most $t$ steps of ${\bf{U_{f01}}}$ and
${\bf{U_{f23}}}$ in parallel, followed by ${\bf{U_{f12}}}$, is,
$$ p_m(t) = \sum_{i = 2}^t p_e(i) $$
In contrast, for non-distributed QPA, the probability of successful completion, after at most $t$ steps, of
${\bf{U}}$,
(the bottom graph of Fig \ref{fg5}) is $P(t) = 1 - (1 - (\al_0\al_1\al_2\al_3)^2 - (\beta_0\beta_1\beta_2\beta_3)^2)^t$.
Figs \ref{fg6} and \ref{fg7} show plots of $p_m(t)$ and $P(t)$ versus $t$ for
$\al_0 = \al_1 = \al_2 = \al_3 = w$ as $w$ varies, and $\al_0 = u, \al_1 = \al_2 = \al_3 = w = 0.9$ as $u$
varies, respectively. For Fig \ref{fg7}, low values of $u$ indicate a contradiction between $x_0$ and the
other three variables. In particular the contradiction is so pronounced when $\al_0 = 0.0$ that successful
QPA completion is highly unlikely. More generally, this indicates that severe internal Factor Graph
contradictions are fatal to QPA (as they are for SPA).
Both Fig \ref{fg6} and \ref{fg7} indicate that, due to initial latency of distributed processing,
non-distributed QPA appears marginally faster for the first few steps. However, after a few steps
distributed QPA in general becomes marginally faster.
In fact results are unfairly biased towards the non-distributed
case, as it is assumed that attempts to complete ${\bf{U}}$ and ${\bf{U_{fij}}}$ have the same
space-time-complexity cost, whereas ${\bf{U}}$ is far more costly. Hence, even for this smallest example,
Distributed QPA outperforms non-Distributed QPA.

\begin{figure}[h]
\centerline{
\epsfig{file=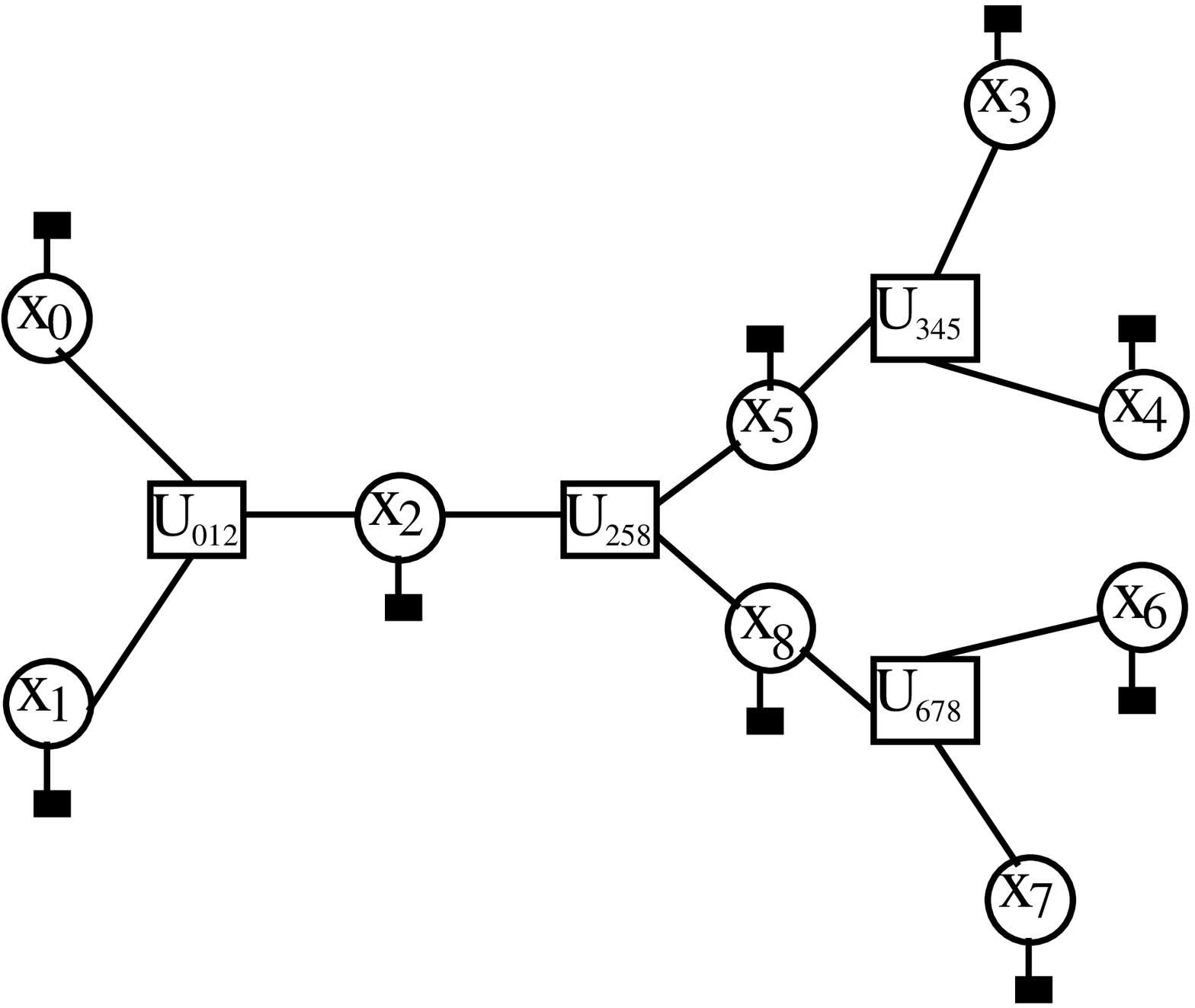,height=10cm}
}
\caption{Distributed QPA, 9-qubits}
\label{fg8}
\end{figure}

The example of Fig \ref{fg5} only achieves marginal advantage using Distributed QPA
because the example has so few nodes. The advantage is more pronounced in Fig \ref{fg8}.

\begin{figure}[h]
\centerline{
\epsfig{file=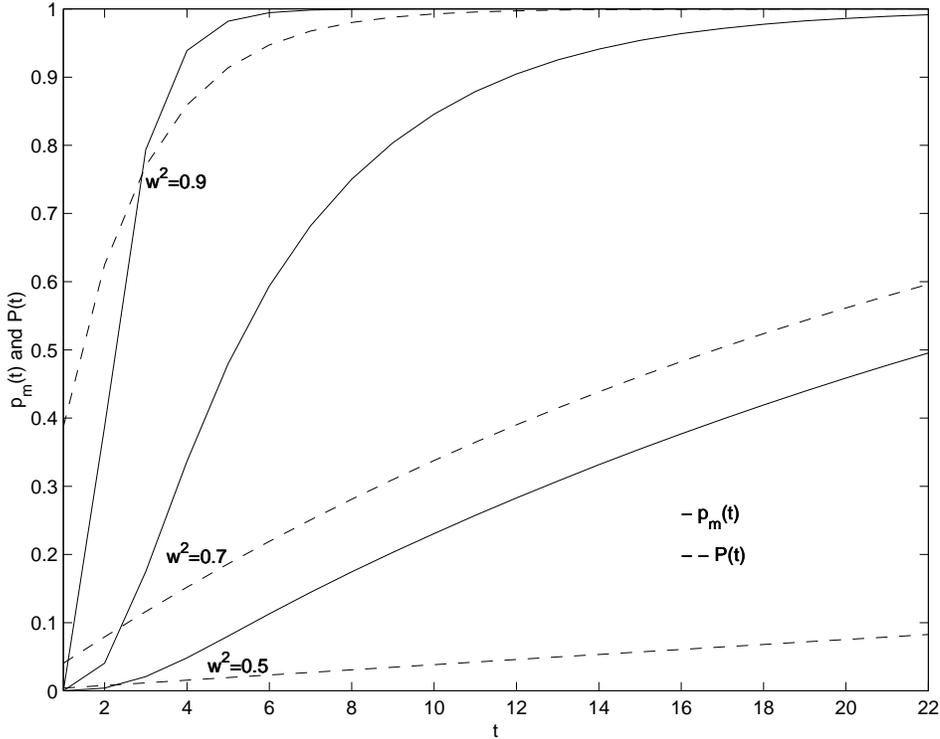,height=10cm}
}
\caption{No of Steps v Non-Distributed and Distributed QPA, 9 qubits}
\label{fg9}
\end{figure}

\begin{figure}[h]
\centerline{
\epsfig{file=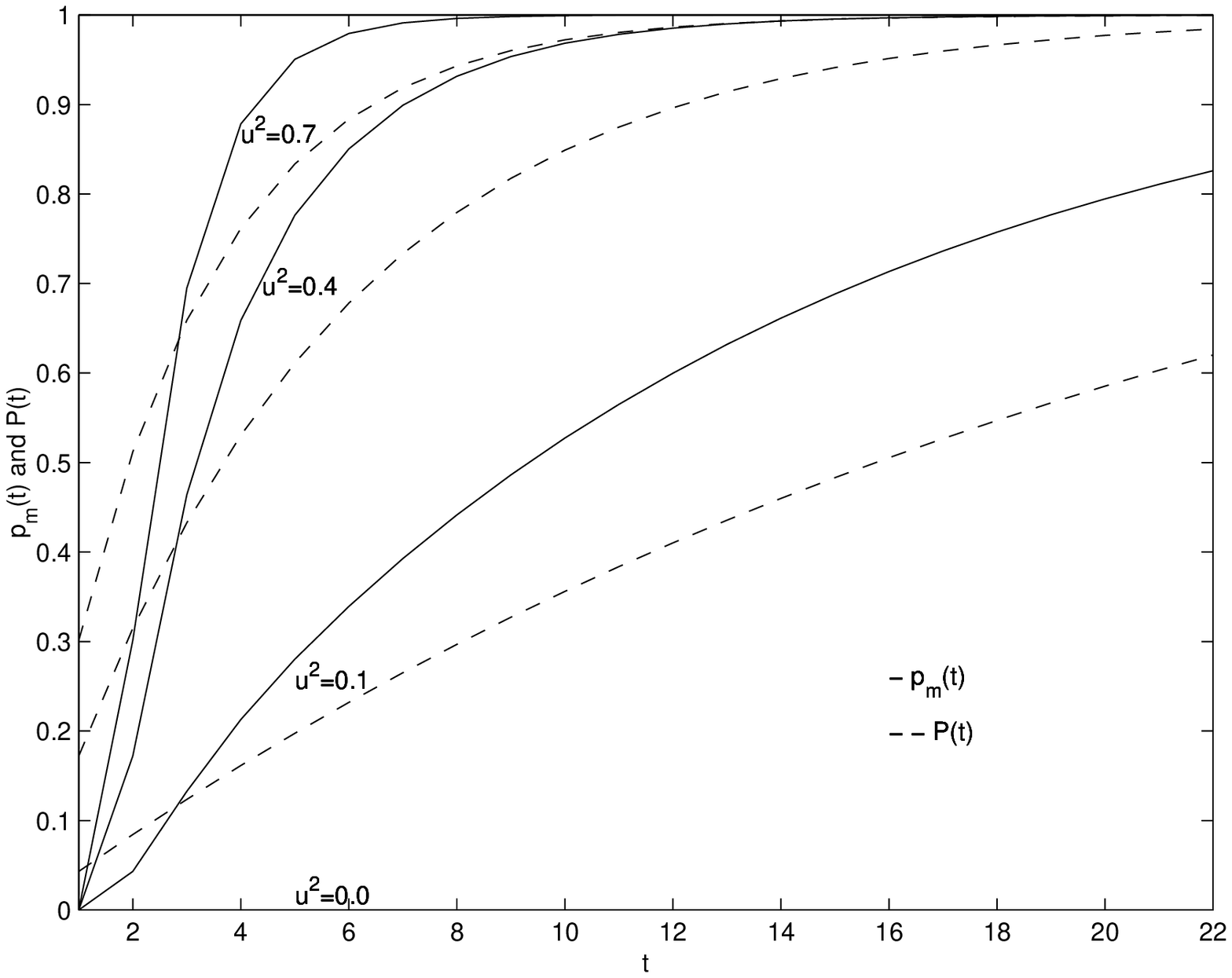,height=10cm}
}
\caption{No of Steps v Non-Distributed and Distributed QPA: $w^2 = 0.9$, $\al_0$ varies, 9 qubits}
\label{fg10}
\end{figure}

Fig \ref{fg8} represents the code ${\bf{C}} = \{000000000,111111111\}$
\begin{footnote}{This code is trivial but demonstrates a 'worst-case' low-rate scenario. In general, codes
of higher rate, with or without cycles, decode more quickly.}
\end{footnote},
where ${\bf{U_{ijk}}} = \m{diag}(1,0,0,1,0,1,1,0)$.
We allow ${\bf{U_{f012}}}$, ${\bf{U_{f345}}}$, and ${\bf{U_{f678}}}$ to operate independently and in parallel.
If ${\bf{U_{f012}}}$ fails to establish, then it does not destroy any successful evolution of
${\bf{U_{f345}}}$ or ${\bf{U_{f678}}}$, as the three localities are not currently entangled. Once
${\bf{U_{f012}}}$, ${\bf{U_{f345}}}$, and ${\bf{U_{f678}}}$ have completed successfully,
the probability of successful subsequent completion of ${\bf{U_{f258}}}$ is, in general, amplified.
Let qubits $x_i, 0 \le i < 9$ of Fig \ref{fg8} initially be in
states $x_i = (\al_i,\beta_i)^T$, where, for notational convenience, we assume all values are real. Then
$\al_i^2 + \beta_i^2 = 1$, $\forall i$.
Let the probability of successful completion after at most $t$ steps
of
 ${\bf{U_{f012}}}$, ${\bf{U_{f345}}}$, and ${\bf{U_{f678}}}$ in parallel, followed by ${\bf{U_{f258}}}$,
be
$p_m(t)$, and the probability of successful completion, after at most $t$ steps, of a non-distributed
version of Fig \ref{fg8} be $P(t)$. Appendix A derives $p_m(t)$ and $P_t$ for this case.
Figs \ref{fg9} and \ref{fg10} show plots of $p_m(t)$ and $P(t)$ versus $t$ for
$\al_i = w$, $\forall i$, as $w$ varies, and $\al_0 = u, \al_i = w = 0.9$, $\forall i$, $i \ne 0$, as $u$
varies, respectively. For Fig \ref{fg10} low values of $u$ indicate contradiction between $x_0$ and the
other eight qubits. The contradiction is so pronounced when $\al_0 = 0.0$ that successful
QPA completion is highly unlikely.

\begin{figure}[h]
\centerline{
\epsfig{file=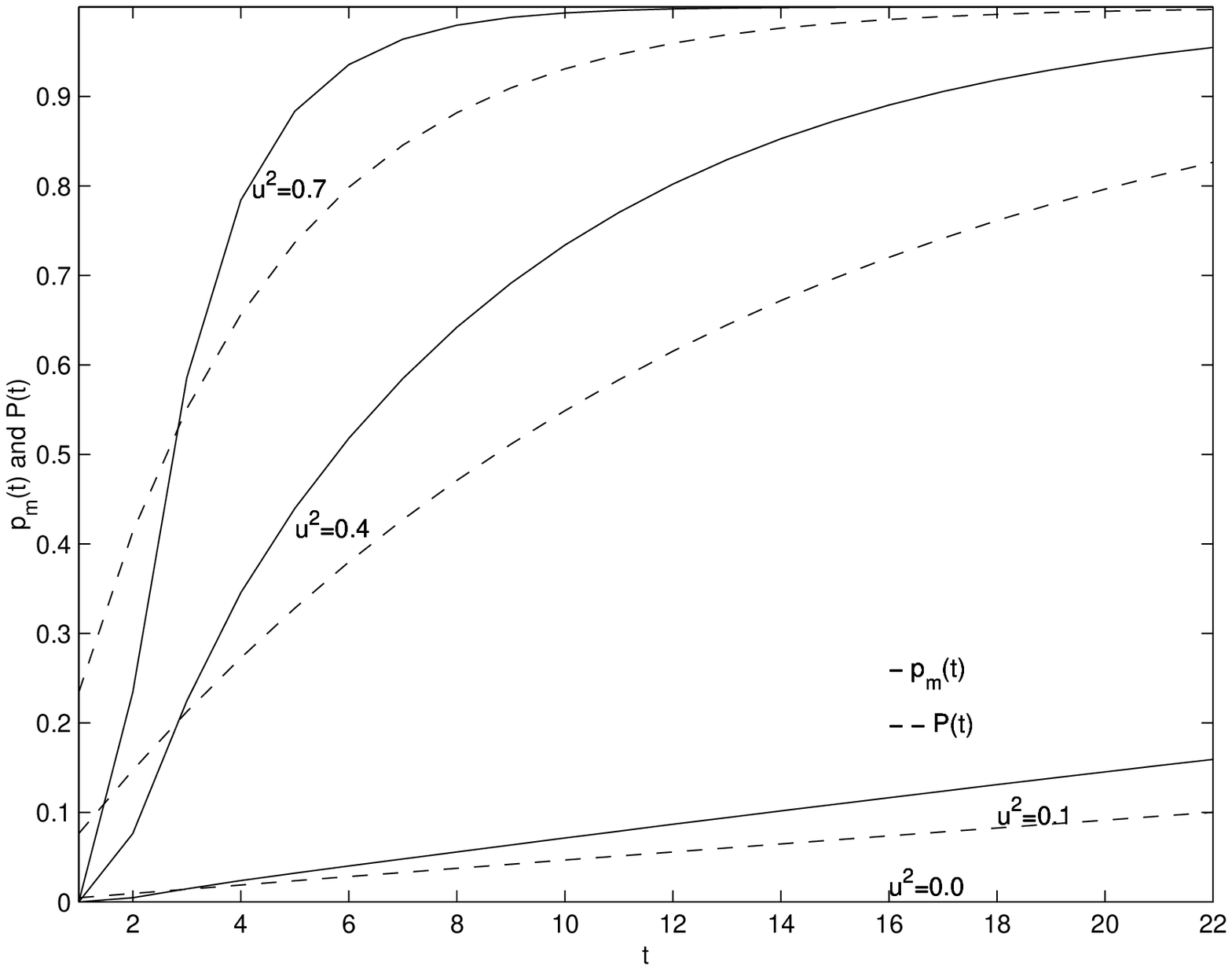,height=10cm}
}
\caption{No of Steps v Non-Distributed and Distributed QPA: $w^2 = 0.9$, $\al_0 = \al_1$ varies, 9 qubits}
\label{fg11}
\end{figure}

\begin{figure}[h]
\centerline{
\epsfig{file=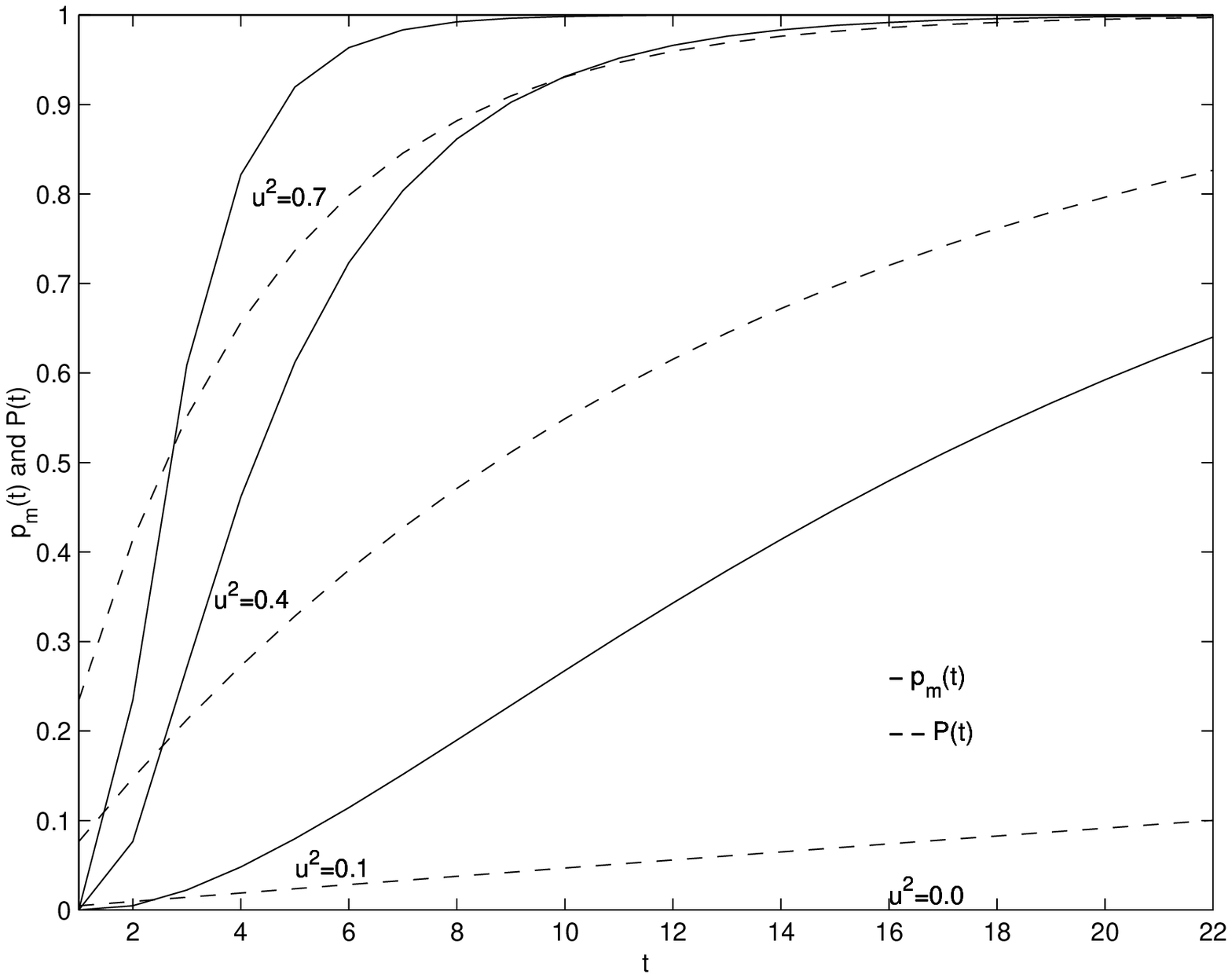,height=10cm}
}
\caption{No of Steps v Non-Distributed and Distributed QPA: $w^2 = 0.9$, $\al_0 = \al_8$ varies, 9 qubits}
\label{fg12}
\end{figure}

Figs \ref{fg11} and \ref{fg12} show plots of $p_m(t)$ and $P(t)$ versus $t$ for
$\al_0 = \al_1 = u, \al_i = w = 0.9$, $\forall i$, $i \ne 0,1$, and
$\al_0 = \al_8 = u, \al_i = w = 0.9$, $\forall i$, $i \ne 0,8$, respectively,
as $u$ varies. Both figures indicate contradictions between two qubits and the rest, but
the scattered nature of contradictions for Fig \ref{fg12} ($x_0$ and $x_8$ are connected to
{\underline{different}} local functions) enhances
Distributed QPA performance compared to Fig \ref{fg11}.

Figs \ref{fg6}-\ref{fg12} indicate that
distributed QPA completes significantly faster than non-distributed QPA, in particular for cases
requiring many steps, $t$.
Even more so as the presented results are unfairly biased towards the non-distributed
case, as it is assumed that attempts to complete non-distributed ${\bf{U}}$ or each constituent
${\bf{U_{fijk}}}$ have the same
space-time-complexity cost, whereas ${\bf{U}}$ is far more costly.
We conclude that Distributed QPA is essential for large Quantum Factor Graphs.
\subsection{Free-Running Distributed QPA}
\begin{figure}[h]
\centerline{
\epsfig{file=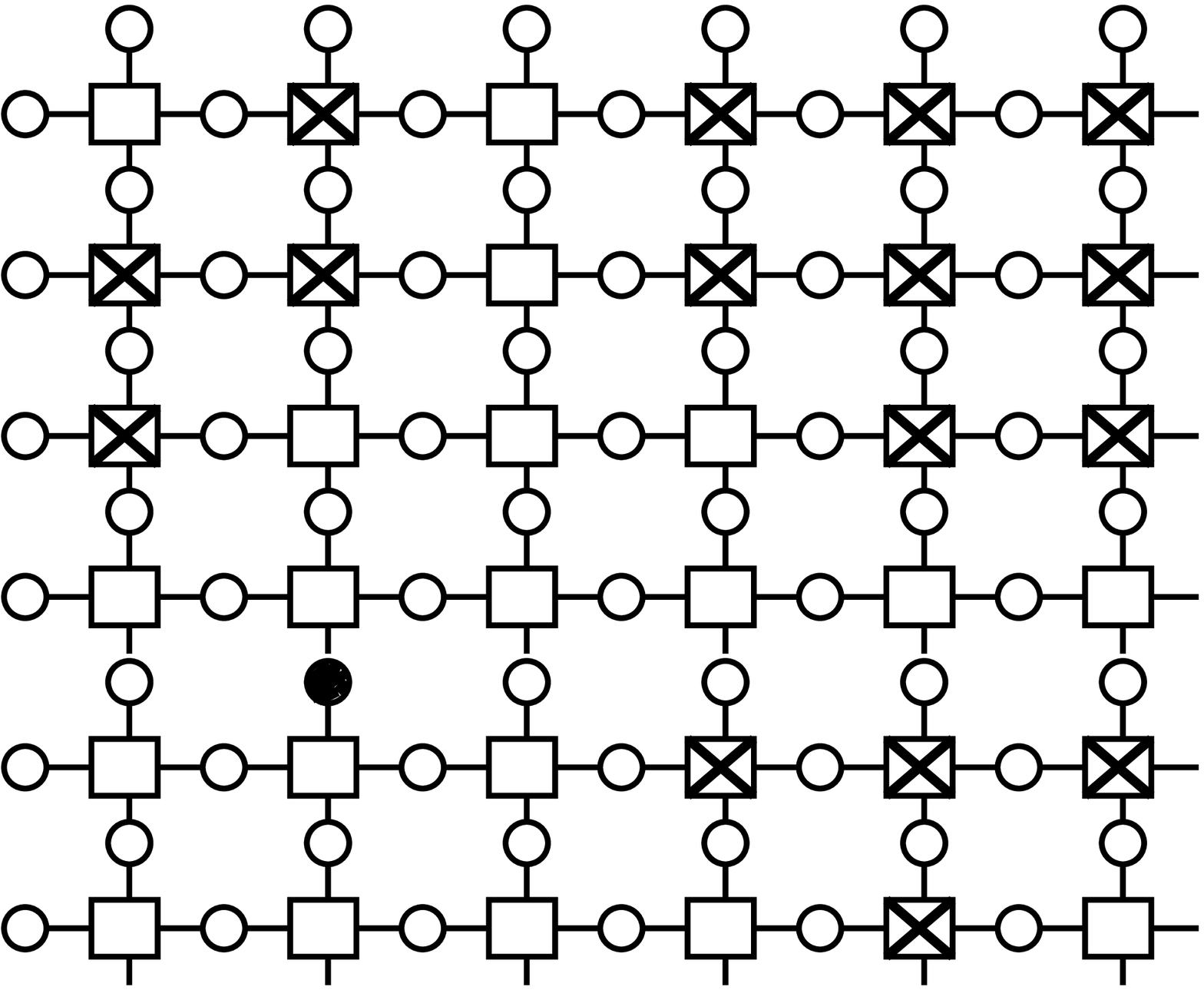,height=7cm}
}
\caption{Free-Running Distributed QPA with one 'Bad' Variable}
\label{fg13}
\end{figure}

Consider the notional Factor Graph of Fig \ref{fg13}. Each (square) function node activates time-independently
on its local (circular) variable nodes. Functions successfully completed are marked with an 'X'.
After a certain time, say, three 'areas of success' evolve, due to general agreement between
input variable states at these localities. This means that variables on the perimeter of each region of success are
'encouraged' to agree with the 'general view' of the associated region of success. Unfortunately, in the bottom
left of the graph is a variable (dark circle) which strongly contradicts with the rest of the graph. No
area of success evolves around it, and it is difficult for other areas of success to 'swallow' it.
Assuming the contradiction is not too strong then, eventually, after numerous attempts, the complete graph
is marked with 'X's and the Graph evolves successfully. At this point the contents of each
qubit variable can be amplified, and final measurement of all qubits provides the ML codeword with high
probability. The advantage of a free-running strategy, where each function node is free to activate
asynchronously,
is that regions of general agreement develop first and influence other areas of the graph to 'follow their
opinion'. Fig \ref{fg13} also shows that one 'bad' (contradictory) qubit can be a fatal stumbling block
to successful evolution of the whole graph (as it can for SPA on classical graphs). Thus Distributed QPA requires Fault-Tolerance, where only an
arbitrary subset of entangled nodes are required as a final result (node redundancy).
The free-running schedule of Fig
\ref{fg13} naturally avoids the 'bad' qubits, but sufficient evolution occurs when enough
function nodes complete. Alternatively, bad qubits could be set to $(\sqrt{0.5},\sqrt{0.5})$ after a
time-out.
A more detailed proposal of Fault-Tolerant QPA is left for future work.

Fig \ref{fg13} also serves to illustrate the 'template' for a Reconfigurable Quantum Graph Array.
One can envisage initialising an array of quantum variables so that two local variables can be strongly
or weakly entangled by identifying the mutual square function nodes with strongly or weakly-entangling
matrices, respectively. In particular, two neighbouring nodes may be 'locally disconnected' by setting the
function node joining them to a tensor-decomposable matrix, (i.e. zero-entangling). The quantum computer
is then measurement-driven. The concept of measurement-driven quantum computation has also recently been
pursued in \cite{Raus:QC}, where a uniform entanglement is set-up throughout the array
\begin{footnote}{It is interesting that this entanglement is strongly related to Rudin-Shapiro
and quadratic constructions \cite{Par:Ent1}} \end{footnote} prior to computation via measurement.

\begin{figure}[h]
\centerline{
\epsfig{file=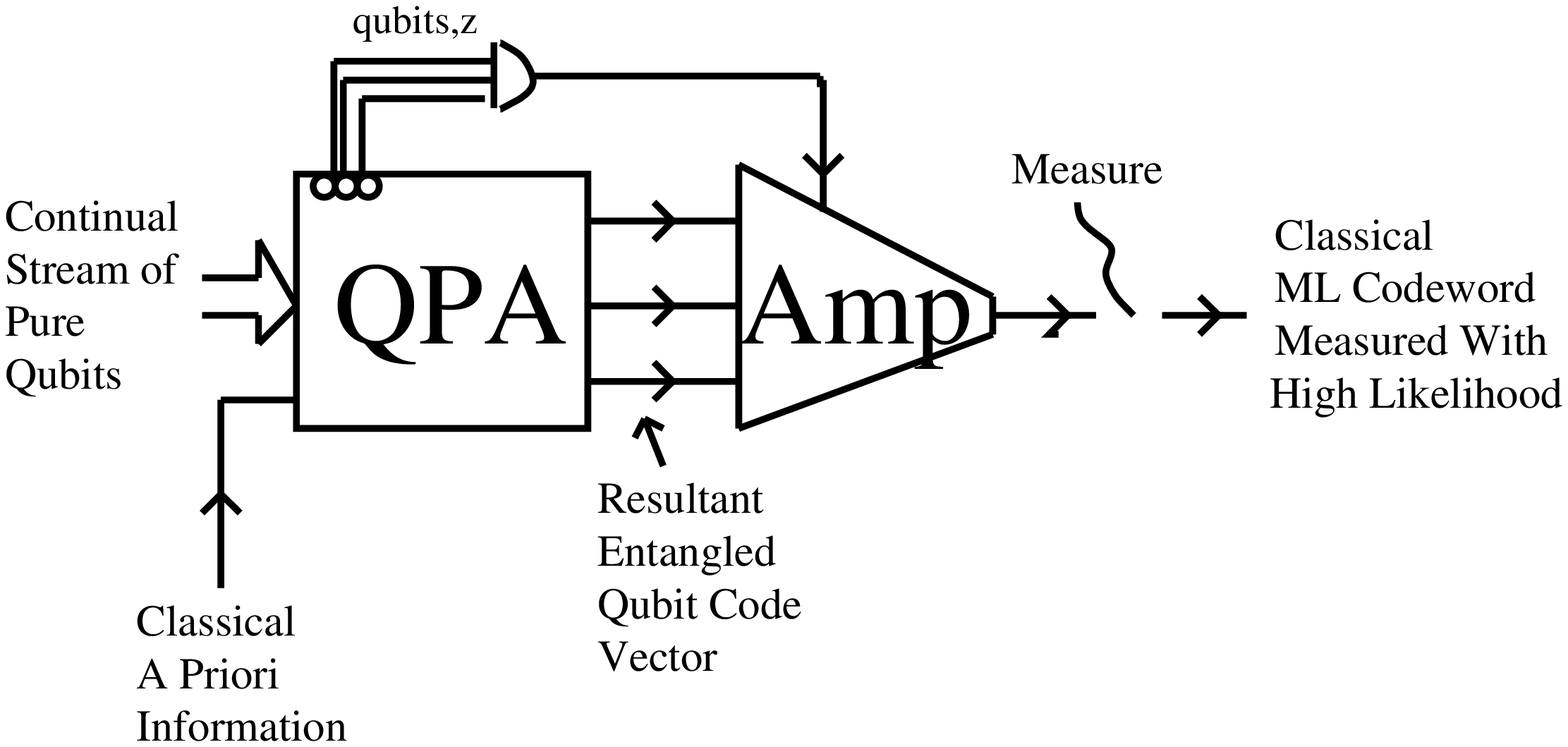,height=6cm}
}
\caption{QPA with Amplification}
\label{fg14}
\end{figure}

Fig \ref{fg14} shows the system view of QPA. A continual stream of pure qubits needs
to be initialised and then entangled, and then amplified, so as to ensure at least one successful entangled
and amplified output from the whole apparatus.

\section{Phase QPA}
\label{sec6}
The above discussions have ignored the capacity of Quantum Systems to carry phase information.
In fact QPA, as presented so far, is immune to phase modification, as classical probabilities have no
phase component. However QPA should be generalised to cope with phase shift
in order to decode quantum information. This is the subject of ongoing research.
\section{Conclusion and Discussion}
\label{sec7}
The Quantum Product Algorithm (QPA) on a Factor Graph has been presented for Maximum-Likelihood (ML) Decoding of
Classical 'soft' information using quantum resources. The relationship of QPA to the Sum-Product Algorithm
(SPA) has been indicated, where
avoidance of summary allows QPA to overcome small graph cycles. Quantum Factor Graphs
use small unitary matrices which each act on only a few
qubits. QPA is measurement-driven and is only statistically likely to
succeed after many attempts. The ML codeword is obtained with maximum likelihood by measuring the entangled
vector resulting from successful QPA. To ensure a high probability of measuring the ML codeword QPA output can be
amplified prior to measurement. The complete ML decoder is only successful after many attempts. Finally,
free-running Distributed QPA is proposed to improve the likelihood of successful QPA completion. The
free-running distributed structure suggests further benefit will be obtained by introducing Fault-Tolerance
in the form of redundant function and variable nodes. Phase aspects of QPA have yet to be
explored. This paper has been written to demonstrate the exponential capacity of quantum systems, and their
natural suitability for graph decompositions such as the Factor Graph. The
paper has not tried to deal with quantum noise and quantum decoherence, but one can
expect the Factor Graph form to 'gracefully' expand to cope with the extra redundancy necessary to protect
qubits from decoherence and noise. When viewed in the context of entangled space, it is surprising how
successful classical message-passing algorithms are, even though they are restricted to operate in
tensor product space. This suggests that methods to improve the likelihood of successful QPA completion may
include the possibility of hybrid QPA/SPA graphs, where SPA operates on non-cyclic and resolvable parts of the graph,
leaving QPA to cope with small cycles or unresolved areas of the graph.
\section{Appendix A - Deriving $p_m(t)$ and $P(t)$ for Fig \ref{fg8}}
The probability of successful completion of ${\bf{U_{f012}}}$, is
$p_{f012} = (\al_0\al_1\al_2)^2 + (\beta_0\beta_1\beta_2)^2$, and similarly for $p_{f345}$
and $p_{f678}$.
Let $h_{012} = (1 - p_{f012})^{q-1}$, $h_{345} = (1 - p_{f345})^{q-1}$, $h_{678} = (1 -
p_{f678})^{q-1}$.
Then the probability of successful completion of ${\bf{U_{f012}}}$, ${\bf{U_{f345}}}$,
and ${\bf{U_{f678}}}$ after exactly $q$ parallel attempts is,

\begin{small}
$$ \begin{array}{c}
p_{0-3-6}(q) = h_{012}h_{345}h_{678}p_{f012}p_{f345}p_{f678} + (1 - h_{012})h_{345}h_{678}p_{f345}p_{f678} \\
 + h_{012}(1 - h_{345})h_{678}p_{f012}p_{f678} + h_{012}h_{345}(1 -
 h_{678})p_{f012}p_{f345} \\
 + (1 - h_{012})(1 - h_{345})h_{678}p_{f678} + (1 - h_{012})h_{345}(1 - h_{678})p_{f345}
 \\
 + h_{012}(1 - h_{345})(1 - h_{678})p_{f012}
\end{array} $$
\end{small}

Given successful completion of ${\bf{U_{f012}}}$, ${\bf{U_{f345}}}$,
and ${\bf{U_{f678}}}$,
the probability of {\bf{subsequent}} successful completion of ${\bf{U_{f258}}}$ is,
$$ p_{f258}' = \frac{(\al_0\al_1\al_2\al_3\al_4\al_5\al_6\al_7\al_8)^2 +
(\beta_0\beta_1\beta_2\beta_3\beta_4\beta_5\beta_6\beta_7\beta_8)^2}{p_{f012}p_{f345}p_{678}} $$
Therefore the probability of successful completion of ${\bf{U_{f012}}}$, ${\bf{U_{f345}}}$,
and ${\bf{U_{f678}}}$,
immediately followed by successful completion of ${\bf{U_{f258}}}$ is,
$ p_{0\rightarrow 8}(q) = p_{0-3-6}(q-1)p_{f258}'$, and the probability of successful completion of
${\bf{U_{f012}}}$, ${\bf{U_{f345}}}$, and ${\bf{U_{f678}}}$, immediately followed by completion
failure of ${\bf{U_{f258}}}$ is,
$ {\overline{p_{0\rightarrow 8}}}(q) = p_{0-3-6}(q-1)(1 - p_{f258}')$.
The probability of successful completion after exactly $t$ steps of ${\bf{U_{f012}}}$, ${\bf{U_{f345}}}$,
and ${\bf{U_{f678}}}$ in parallel, followed by ${\bf{U_{f258}}}$, is,
$$ p_e(t) = \sum_{q=2}^t p_{0\rightarrow 8}(q) \sum_{{\bf{v \in D(t-q)}}} \prod_{u \in {\bf{v}}}
{\overline{p_{0\rightarrow 8}}}(u) $$ where ${\bf{D(k)}}$ is the set of unordered partitions of
$k$. Therefore the probability of successful completion after at most $t$ steps of
${\bf{U_{f012}}}$, ${\bf{U_{f345}}}$, and ${\bf{U_{f678}}}$ in parallel, followed by ${\bf{U_{f258}}}$, is,
$$ p_m(t) = \sum_{i = 2}^t p_e(i) $$
In contrast, the probability of successful completion, after at most $t$ steps, of a non-distributed
version of Fig \ref{fg8} is
$P(t) = 1 - (1 - (\al_0\al_1\al_2\al_3\al_4\al_5\al_6\al_7\al_8)^2 -
(\beta_0\beta_1\beta_2\beta_3\beta_4\beta_5\beta_6\beta_7\beta_8)^2)^t$.

\end{document}